\pdfoutput=1
\documentclass[conference]{IEEEtran}
\IEEEoverridecommandlockouts
\usepackage{xurl}
\usepackage{xcolor}
\definecolor{revgray}{RGB}{0,0,0} % 关闭高亮：revgray == black
\newcommand{\rev}[1]{\textcolor{revgray}{#1}} % 你也可以改成 red/magenta

\usepackage{cite}
\usepackage{amsmath,amssymb,amsfonts}
\usepackage{graphicx}
\usepackage{textcomp}
\usepackage{multirow}
\def\BibTeX{{\rm B\kern-.05em{\sc i\kern-.025em b}\kern-.08em
    T\kern-.1667em\lower.7ex\hbox{E}\kern-.125emX}}

\usepackage{graphicx}
\usepackage{subcaption} % For subfigure captions
\usepackage{comment}

\usepackage{tabularx}
\usepackage{booktabs}
\usepackage{ragged2e}
\newcolumntype{Y}{>{\RaggedRight\arraybackslash}X}

\usepackage{algorithm}
\usepackage{algpseudocode}

\usepackage{enumitem}

\setlist[itemize]{leftmargin=1.2em, itemsep=2pt, topsep=2pt}
\setlist[enumerate]{leftmargin=1.2em, itemsep=2pt, topsep=2pt}

\setlist[enumerate,1]{label=(\arabic*), ref=\arabic*}
\setlist[enumerate,2]{label=(\arabic*), ref=\arabic*}

\begin{document}

\title{GALA: Generative Aligned Learning for Adaptive Multimodal Representation in the Taobao Shangou Recommender System}
\author{
\IEEEauthorblockN{
Jiping Liu, Zhongmin Zhang, Zisen Sang\textsuperscript{*}, Zhijia Fang\textsuperscript{*}
}
\IEEEauthorblockA{
\textit{Rajax Network Technology (Taobao Shangou of Alibaba)}\\
Shanghai \& Beijing, China\\
{\textit \{ljp178097, zhongminzhang.zzm, zisen.szs, zhijia.fzj\}@alibaba-inc.com}
\thanks{\textsuperscript{*}Corresponding author.}
}

\vspace{0.6em}

\IEEEauthorblockN{
Tao Ouyang
}
\IEEEauthorblockA{
\textit{Central South University}\\
Changsha, China\\
{ouyangtao@csu.edu.cn}
}

\vspace{0.6em}

\IEEEauthorblockN{
Ma Jiang, Shaopeng Liang, Zeyang Hou, Guodong Cao, Jia Jia
}
\IEEEauthorblockA{
\textit{Rajax Network Technology (Taobao Shangou of Alibaba)}\\
Shanghai \& Beijing, China\\
{\textit\{majiang.mj, shaopengliang.lsp, houzeyang.hzy, guodong.cao, jj229618\}@alibaba-inc.com}
}
}

\maketitle

\begin{abstract}
Modern recommender systems in food delivery increasingly leverage multimodal signals---including images, text, and user interaction histories---to enhance user experience, yet effective fusion of these heterogeneous modalities remains challenging due to discrepancies in semantics, scale, and update frequency, hindering both the joint modeling of multimodal signals and adaptation to evolving user intent. In mainstream two-stage approaches, the separation between content-semantic pretraining of image--text encoders and behavior-driven ranking models limits alignment between semantic understanding and user behavior patterns. To address these issues, we present GALA, a three-stage pipeline extending existing learning frameworks, whose core innovation lies in an intermediate ``generative RL alignment'' stage that constructs multimodal pretraining data from user behavior and refines it via conversion-based rewards, effectively bridging the pretraining--fine-tuning gap to align with downstream objectives. Specifically, GALA comprises three tightly coupled stages: first, behavior-aware triplet pretraining on query--image--text pairs from search logs to early capture user intent and content preferences; second, the novel intermediate stage, which refines multimodal embeddings through reward-driven optimization (GRPO) to dynamically align them with user behavior and bridge the pretraining--fine-tuning gap; and finally, integration of multimodal and ID embeddings via adaptive gating with a hybrid loss, preserving multimodal contributions under long-term ID-dominant training. GALA has been deployed in the production environment at Taobao Shangou, serving over 200 million daily active users. Compared with state-of-the-art (SOTA) methods, it delivers consistent offline gains of +0.12/+0.20 AUC along with better PCOC metrics. Large-scale online A/B tests further report a 0.55\% increase in order volume, confirming GALA’s effectiveness at industrial scale and its robustness across diverse demand patterns.
\end{abstract}


\section{Introduction}
\begin{figure}[t]
  \centering
  \includegraphics[width=\linewidth]{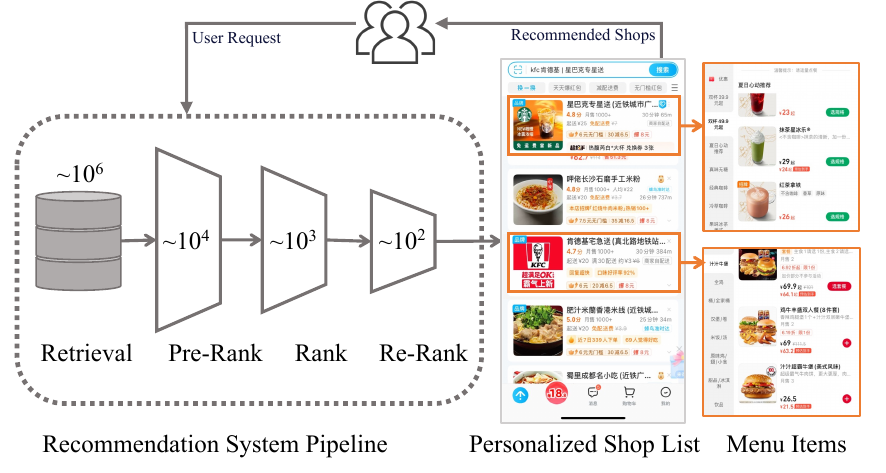}
  \caption{Taobao Shangou Recommender System Architecture.}
  \label{fig1}
\end{figure}

Taobao Shangou is a leading food delivery platform that ensures efficient service to its customers. Its recommender system serves hundreds of millions of users, millions of shops, and billions of items, playing a pivotal role in user engagement. As illustrated in Fig.~\ref{fig1}, when a user submits a request, the system first retrieves a set of deliverable shops. These candidates undergo a cascaded procedure before a ranked list is presented to the user. Upon clicking on the preferred shops, the user proceeds to place an order, which will be finally delivered to a specified address.

Multimodal features---particularly shop images and textual descriptions---play a critical role in user decision-making within recommender systems. While two shops may offer similar cuisines (e.g., burger restaurants), their visual presentations and menu descriptions often differ substantially, directly influencing user preferences beyond mere dietary alignment. Traditional ID-based ranking models~\cite{agr:19, ps:22, lsy22, jgy23, jzh25, lyd:2024}, which derive embeddings exclusively from interaction histories, fail to capture these decisive multimodal characteristics. This limitation results in information loss, particularly for new and long-tail shops with sparse interaction data. Consequently, these shops remain inadequately modeled and suffer reduced exposure, ultimately compromising overall recommendation quality and platform performance.

\rev{
In large-scale industrial recommender systems, existing multimodal-enhanced ranking solutions can generally be categorized into two paradigms. \textbf{End-to-end joint training}~\cite{xlx:2024, xhq:2025} updates multimodal content encoders and ID/behavioral representations jointly under the ranking objective (e.g., Click-Through Rate (CTR) or Conversion Rate (CVR)), enabling objective-level alignment between representation learning and downstream optimization, often yielding strong offline improvements. 
However, this paradigm is hard to deploy in large-scale production systems. Jointly training/updating large multimodal encoders with ranking models typically requires either per-request multimodal encoding or frequent encoder refresh, both of which conflict with strict \textbf{millisecond-level latency} and high-throughput serving requirements. Moreover, many end-to-end designs rely on cached or memory-bank item representations to reduce online cost~\cite{xlx:2024, xhq:2025}; in fast-evolving, extremely long-tailed inventories, such representations are inevitably refreshed with delay, resulting in \textbf{insufficient coverage} for newly added/updated or long-tail candidates. Therefore, while algorithmically appealing, end-to-end approaches are not directly applicable under our constraints. In our production setting, the serving stack follows the standard paradigm of \emph{offline embedding computation, online key--value (KV) lookup, lightweight ranking}. Under this paradigm, the prevailing solution is the \textbf{two-stage pipeline}~\cite{yl:24, xr:24, wmk:2024, lty:2024}, where a multimodal encoder is first pretrained on large-scale image--text alignment tasks and used to compute (often frozen) item embeddings offline; these embeddings are then retrieved via online KV lookup and fed into a downstream ranker optimized for behavioral objectives such as CTR/CVR. This paradigm is simple to implement and cost-effective to iterate, but the decoupling between multimodal representation learning and ranking optimization often leads to misalignment, limiting the actual contribution of multimodal features to final ranking performance.
} 

In light of these issues, this work identifies three principal challenges: \textbf{(i) Objective and Distribution Misalignment.} A fundamental discrepancy exists between the static, content-driven distribution of the pretraining phase and the dynamic, behavior-driven objective of the ranking stage. The pretrained encoders produce static, content-oriented embeddings, whereas the ranking model must adapt to evolving user intent. When embeddings are frozen at serving time, they cannot reflect recent interaction data, thus limiting the contribution of multimodal representations to ranking performance. \textbf{(ii) Domain-Specific Representation Gap.} Effective multimodal learning must capture nuanced, domain-specific factors that influence user decisions. In food delivery, for instance, user preferences are highly contextual: key dish components (e.g., meats) often outweigh staples (e.g., rice) in importance, and preferences vary significantly by occasion (e.g., breakfast vs. afternoon tea) or taste profile (e.g., a preference for spicy cuisines). General-purpose multimodal pretraining lacks the targeted adaptation needed to distill these fine-grained, domain-specific features~\cite{vdo:17,cw:21,sr:23,jc:24}. \textbf{(iii) Ineffective Fusion in ID-Dominant Architectures.} On platforms with extensive historical data, ranking models are typically dominated by long-term ID-based signals (e.g., user, item, shop IDs). Simply incorporating multimodal features into these established models yields limited gains, as the influence of multimodal signals is suppressed by the strong ID-centric baselines. A dedicated fusion mechanism is therefore necessary to ensure multimodal features contribute meaningfully.

Taken together, these challenges highlight the need for a multimodal food recommender system that can couple fine-grained content understanding with dynamic user intent modeling, forming a cognitive feedback loop that adapts to evolving preferences. Therefore, we propose \textbf{GALA}, a three-stage pipeline extending existing learning frameworks, whose core innovation lies in an intermediate ``generative RL alignment'' stage that constructs multimodal behavior-alignment data from user behavior and refines it via conversion-based rewards (derived from purchase events), thereby bridging the pretraining--fine-tuning gap to align with downstream objectives. Specifically, our \textbf{GALA} comprises: (i) domain-specific pretraining of multimodal encoders to capture culinary characteristics, (ii) generative behavior alignment to iteratively refine multimodal representations through historical interaction signals, and (iii) adaptive fusion of multimodal and ID features through a novel gating mechanism. As a new post-training stage, the generative behavior alignment leverages reinforcement learning to refine image--text embeddings by capturing evolving user preferences. On this basis, the adaptive fusion module combines these frozen, enhanced embeddings with trainable ID embeddings, efficiently balancing contribution trade-off among diverse features at the application stage.

To summarize, our contributions are as follows:

\textbf{Domain-Adaptive Cross-Modal Alignment}: We propose a domain-adaptive cross-modal alignment framework which learns intent-aware representations by optimizing a contrastive loss over query-shop and query-content triplets mined from large-scale search logs, explicitly capturing how subtle content differences shape user preference at the pretraining stage.

\textbf{Generative User Behavior Alignment}: To overcome the static nature of conventional multimodal embeddings, we develop a generative, behavior-aligned refinement procedure, which employs Group Relative Policy Optimization (GRPO) to update embeddings via reward-driven optimization with an auxiliary objective, enriching their expression of personalized intents during the post-training stage.

\textbf{Multimodal Embedding Fusion for RecSys}: Recognizing the dominance of ID features in industry-scale systems, we design a novel adaptive gating mechanism to seamlessly integrate multimodal and ID embeddings at the application stage. Moreover, we design a hybrid loss, a primary loss on the fused representation and an auxiliary loss on the multimodal branch, enabling the model to balance these signals in a data-driven manner.

 Our \textbf{GALA} demonstrates superior performance compared to state-of-the-art (SOTA) methods in offline evaluations. Furthermore, online A/B testing on Taobao Shangou platform confirms its practical effectiveness, yielding a statistically significant 0.55\% increase in total order volume.

 \begin{figure*}[t]
  \centering
  \includegraphics[width=0.8\linewidth]{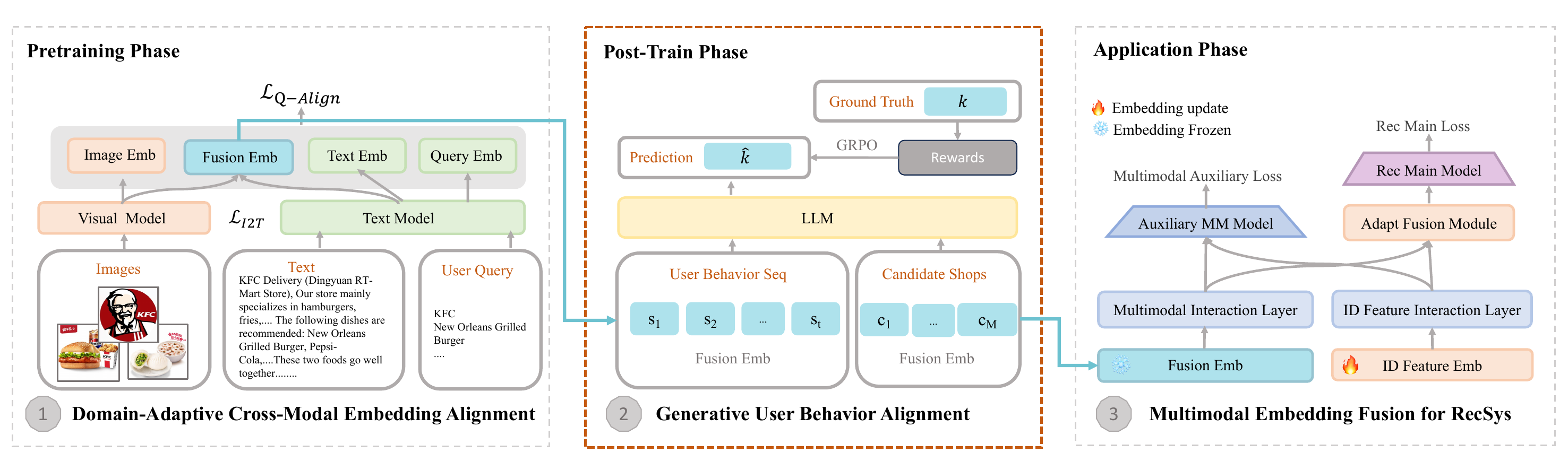}
  \caption{Overview of the Three-Stage Framework \textbf{GALA}.}
  \label{fig2}
\end{figure*}

\section{Preliminaries}
Before delving into our method, we introduce basic knowledge of the industrial food delivery recommender system: \textbf{(i) System Objective}: Generate personalized and efficient shop ranking lists by learning from massive real-time user--shop interaction data (i.e., labels) and discrete ID-based features (i.e., model inputs). The ranking model optimizes accurate predictions of user preferences from these signals. A high-quality ranking list places shops that users are likely to be interested in the top positions, thereby improving key metrics such as CTR and conversion rate. \textbf{(ii) Feature Taxonomy}: Modern ranking models typically employ three categories: \textbf{1) ID-based features} describe user profiles, the target shop (to be predicted), and the contextual information; \textbf{2) User behavior sequences} describe user historical interacted items; \textbf{3) Multimodal representation features} describe items' text/visual information. \textbf{(iii) Model Architecture} The standard ID-based ranking model typically adopts an embedding and Multi-Layer Perceptron (MLP) architecture. In this way, categorical ID features are first transformed into dense embeddings via an embedding layer. To capture user interests, a historical behavior modeling module assesses the relevance between the target item’s embedding and the embeddings of the user’s historical interactions. This module generates fixed-length representation vectors by aggregating the target item embedding with the user’s interaction history. These aggregated representations are then concatenated with other ID embeddings to form the input for an MLP. Finally, the MLP processes this combined input to produce the prediction score.

\section{Related Work}
\label{sec:related_work}

\textbf{Multimodal Alignment via Vision-Language Pretraining:} Recent work leverages vision-language pretraining to align image and text into a unified semantic space. For instance,
\cite{ar:21} introduced a natural language-supervised framework for cross-modal alignment. \cite{ay:22} developed Chinese CLIP, which is pretrained on 200 million Chinese image--text pairs. \cite{jl:21} proposed ALBEF, combining contrastive alignment with multimodal fusion via co-attention and momentum distillation. More recently, \cite{xz:24} presented GME, a large multimodal embedder using synthetic data and contrastive learning, achieving SOTA on UMRB. While these methods excel at semantic alignment, they generate generic embeddings that are not tailored for recommender systems.

\textbf{Multimodal Fusion in Recommender Systems:} Several studies integrate multimodal features into recommendation pipelines. \cite{cw:21} extracted image regions via Mask-RCNN and fused them with text using co-attention. \cite{yl:24} aligned BEiT-3 embeddings with ID-based collaborative signals. \cite{xr:24} designed a two-phase framework: semantic pretraining followed by industrial deployment via SimTier/MAKE. \cite{by:25} combined generative pretraining with DLRMs, enabling scalable deployment via simple similarity matching. However, these methods treat multimodal embeddings as static inputs, lacking adaptive fusion mechanisms to dynamically adjust visual and textual feature weights based on user preferences. Moreover, these approaches assume clean, single-image inputs---an unrealistic assumption for real-world food delivery platforms, where dishes often have multiple ambiguous, redundant, or low-quality images.
\rev{Moreover, \cite{xlx:2024} proposed EM3, which fine-tunes content representations within the ranking model and adopts Low-Rank Adaptation (LoRA) to reduce the training cost of modeling long multimodal user sequences. \cite{xhq:2025} proposed LEMUR, which trains content and ID representations end-to-end and introduces a memory-bank mechanism to alleviate the computational bottleneck of multimodal sequential modeling. However, fine-tuning or memory-bank based end-to-end designs may suffer from \textbf{insufficient coverage} for newly added/updated and long-tail candidates when representations are refreshed only after training, which can be particularly problematic in fast-evolving, long-tailed inventories.}

\textbf{Discrete Semantic Representations.} Discrete semantic IDs have been explored to improve generalization by clustering similar items and enabling code-based generative retrieval in sequential recommendation (e.g., via VQ/RQ-VAE quantization and autoregressive code prediction)~\cite{sr:23,bz:24}. While effective for sequence generation and retrieval, these approaches provide limited fine-grained cross-modal interaction and are less compatible with end-to-end relevance ranking in our industrial setting; thus, they are not the focus of this paper.

\section{Methodology}
\subsection{Framework Overview}
As an extension to conventional approaches, we introduce \textbf{GALA}, a domain-adaptive, generatively-aligned framework comprising three stages tailored for food delivery recommendation, as illustrated in Fig.~\ref{fig2}:

\textbf{Stage 1} (\emph{Domain-Adaptive Cross-Modal Alignment}), GALA constructs domain-specific query--image--text triplets from large-scale user search logs, and uses contrastive learning to align the multimodal embeddings. This early-stage alignment captures user intent signals and builds user-centric content representations grounded in real interaction data.

\textbf{Stage 2} (\emph{Generative User Behavior Alignment}), the framework adopts a generative reinforcement learning paradigm---specifically, next-shop prediction---directly on user behavior sequences. Conversion-based rewards drive the continuous optimization of multimodal embeddings, bridging the pretraining--fine-tuning gap and ensuring inherent alignment with downstream business objectives.

\textbf{Stage 3} (\emph{Multimodal Embedding Fusion for RecSys}) introduces an adaptive gating mechanism to dynamically balance the contributions of ID-based and multimodal representations. An auxiliary supervision term regulates the gate weights to preserve multimodal effectiveness under long-term ID-dominant training, mitigating performance degradation.

Together, these stages enable GALA to unify multimodal pretraining, behavior-driven alignment, and adaptive fusion in a unified three-stage recommendation pipeline.

\subsection{Domain-Adaptive Cross-Modal Alignment}

Effective multimodal representations for food delivery recommendations must simultaneously encode domain-specific semantics and user intent patterns.  Unlike generic domains, food platforms exhibit two critical characteristics: (i) Highly localized culinary terminology (e.g., ``ants climbing trees'' refers to pork vermicelli, not literal insects or trees), which generic multimodal models fail to interpret, and (ii) Focused user attention on specific visual/textual attributes during decision-making. To address these issues, we specifically design dual alignments as follows:

\subsubsection{Domain-Specific Image--Text Alignment}
We first construct a training dataset using domain-specific signals to bridge the semantic gap in food delivery terminology. This dataset enables models to accurately interpret localized dish names and categories---a critical capability where generic models fail to associate colloquial terms (e.g., ``ma yi shang shu'' [ants climbing trees]) with their actual culinary referents (pork vermicelli). Specifically, positive image--text pairs are derived from structured shop data (e.g., official image--text pairs of dishes/shops), while negative pairs are generated via in-batch negative sampling using Momentum Contrast~\cite{kh:20}, to mitigate pseudo-negative noise.

\begin{figure}[t]
  \centering
  \includegraphics[width=0.65\linewidth]{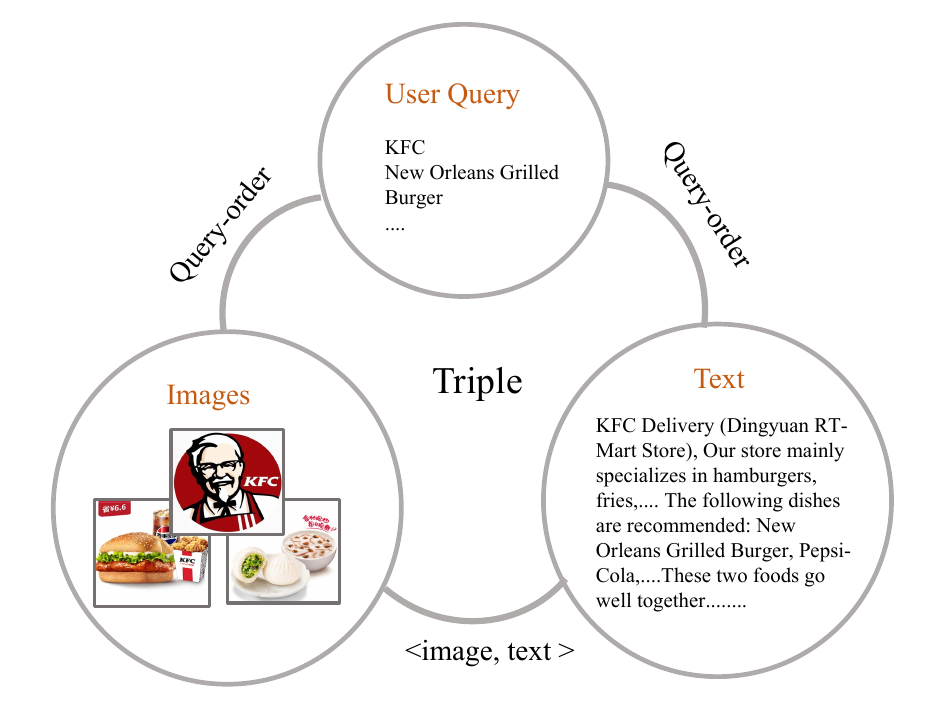}
  \caption{Triplet Data for Domain-Adaptive Cross-Modal Alignment.}
  \label {fig:triple-data}
\end{figure}

In our setting, semantic similarity differences can be naturally represented through users’ positive and negative interactions. When a user’s query in the search scenario leads to a purchase, its semantic correlation with the target shop is markedly stronger compared to non-converted samples. This provides an effective supervision signal for contrastive learning. Accordingly, positive samples are defined as the triplet of (\emph{query}, \emph{image}, \emph{text}) corresponding to the shop where the user ultimately placed the order. Negative samples are taken as other shops within the current mini-batch, serving as potential negatives.
In practice, we construct the candidate training set by mining user search-to-purchase behaviors, extracting the top-10 queries and the associated top-5 purchased items for each shop (based on transaction logs, with non-primary category items excluded). Each (\emph{query}, \emph{image}, \emph{text}) triplet is thus obtained from these high-confidence interactions, as illustrated in Fig.~\ref{fig:triple-data}.

\begin{figure}[t]
  \centering
  \includegraphics[width=0.75\linewidth,height=0.28\textheight,keepaspectratio]{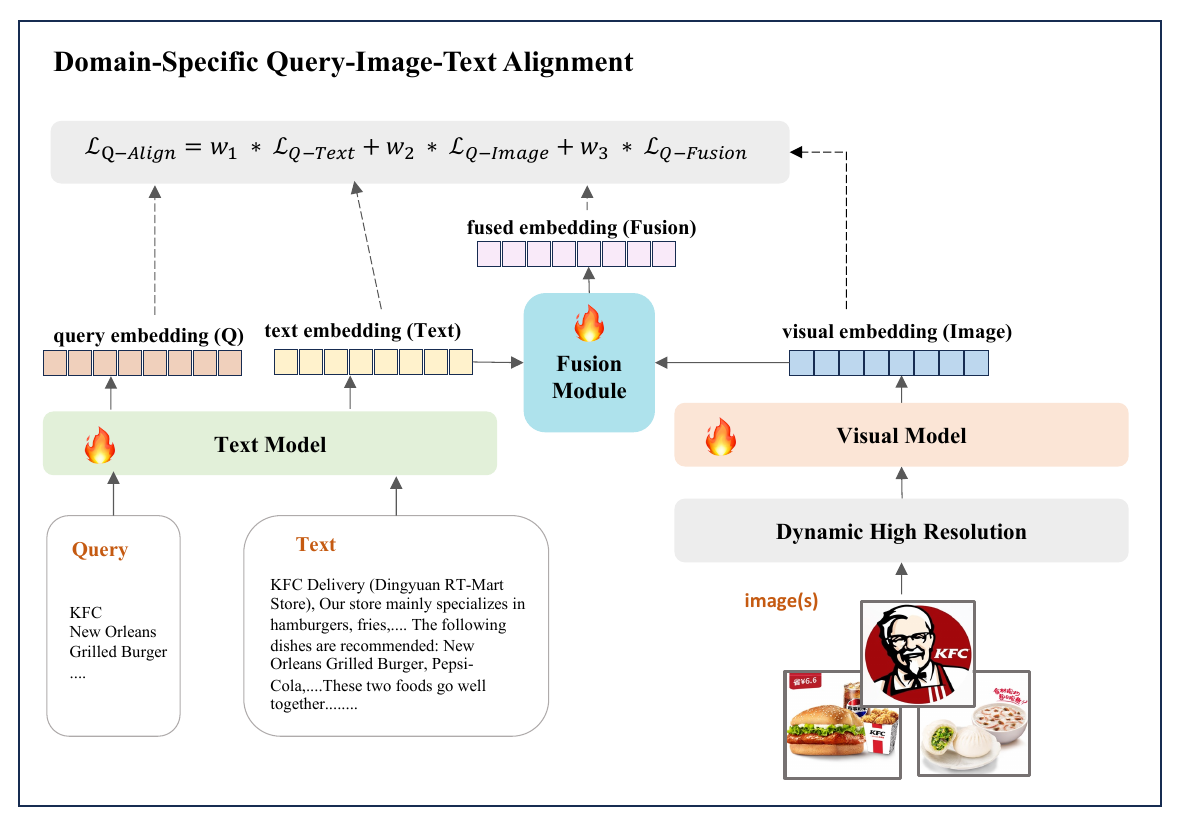}
  \caption{Pretraining Stage of \textbf{GALA}.}
  \label{fig:stage1-train}
\end{figure}

Then we employ two encoders: a hierarchical visual encoder for fine-grained visual feature extraction (e.g., dish colors, shop logo) and a textual encoder for semantic information capture (e.g., menu descriptions, promotional texts). To achieve domain-specific alignment, we leverage contrastive learning to maximize similarity between matched image--text pairs (e.g., a burger image and its corresponding ``juicy beef burger with lettuce'' description) while minimizing similarity for mismatched pairs (e.g., a sushi image paired with an unrelated burger description). This objective ensures domain-consistent visual and textual embeddings. Formally, the contrastive learning based loss function \(\mathcal{L}_{\text{I2T}}\) \cite{ao:18} is defined as:
\begin{equation}
\label{eq1}
\mathcal{L}_{\text{I2T}} = -\log \frac{ \exp(t \cdot i_+) / \tau }{ \sum_{j=0}^{K} \exp(t \cdot i_j) / \tau },
\end{equation}
where \(t\) denotes the text embedding, capturing information from textual descriptions, and \(i\) represents the image embedding, encoding visual features. Specifically, \(i_+\) signifies the positive sample, where the image embedding \(i_+\) is semantically aligned with the text \(t\), while \(i_j\) (where \(j = 0, 1, \dots, K\)) are negative samples, which are image embeddings that do not match the text context. This objective maximizes the log-probability of the positive sample \(i_+\) against a set of negative samples \(i_j\) under the temperature parameter \(\tau\), thereby learning consistent multimodal representations.

\subsubsection{Query-Shop Alignment with Multi-Loss Optimization}
After domain alignment, we further align user queries (reflecting intent) with shop multimodal embeddings as shown in Fig.~\ref{fig:stage1-train}. Query-alignment positive pairs are built from user behavior: queries paired with shops users purchased from (stronger relevance). Negatives are selected from in-batch non-purchased candidates, where batches are custom-sampled using domain knowledge (e.g., shop categories, cuisine types) to provide clear supervisory signals for the triplet loss. Thus, the following complementary loss functions should be jointly optimized with weighted contributions to refine shop embeddings for precise intent-aware alignment:

\textbf{Query-text loss.} Aligns query semantics with shop textual embeddings (e.g., matching ``spicy'' with a shop’s ``Sichuan hotpot, numbing-spicy flavor'' description) via contrastive learning, enhancing semantic consistency between user intent and textual features; 

\textbf{Query-image loss.} Ensures query intent matches shop visual embeddings (e.g., aligning ``bright, lively dining environment'' with a shop’s colorful, bustling interior images), strengthening consistency between intent and visual features; 

\textbf{Query-fusion loss.} Focuses on the fused embedding of a shop (integrating visual and textual features), using contrastive learning to align queries with this comprehensive representation---ensuring user intent matches the holistic impression of the shop. 

We employ late fusion---processing modalities separately before integration, since it outperforms intermediate interaction by preserving fine-grained visual details critical for dish recognition (e.g., identifying cilantro in ingredients), consistent with NoteLLM-2 \cite{cz:25}. Additionally, we conduct a comparative analysis:  MLP and T5 \cite{cr:20} fusion mechanisms. Formally, the loss function \(\mathcal{L}_{\text{Q-Align}}\) is defined as:
\begin{equation}
\label{eq2}
\mathcal{L}_{\text{Q-Align}} = w_{\text{1}} \mathcal{L}_{\text{Q-Text}} 
+ w_{\text{2}}\mathcal{L}_{\text{Q-Image}} 
+ w_{\text{3}} \mathcal{L}_{\text{Q-Fusion}}.
\end{equation}
All losses follow the contrastive formulation of \(\mathcal{L}_{\text{I2T}}\), differing only in the aligned representations. The weights \(w_{\text{1}}, w_{\text{2}}, w_{\text{3}}\) balance different modal-alignment losses, adapting the model to downstream tasks. \(Q\) stands for the user query, representing the intent or search term input, \(Text\) and \(Image\) represent the textual and visual embeddings of a shop, respectively. The fused representation, denoted as \(Fusion\), combines these multimodal features to produce a comprehensive shop embedding.

\begin{figure}[t]
  \centering
  \includegraphics[width=0.75\linewidth,height=0.28\textheight,keepaspectratio]{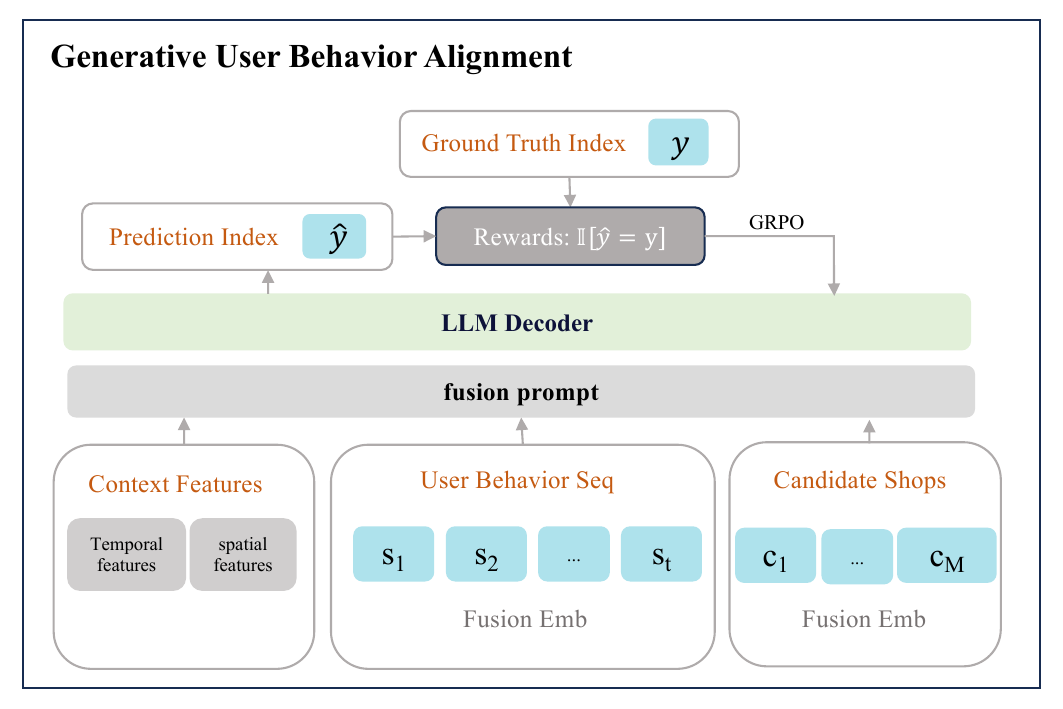}
  \caption{Post-Training Stage of \textbf{GALA}.}
  \label{fig:stage2-train}
\end{figure}

{\color{revgray}
\subsection{Generative User Behavior Alignment}
\label{sec:stage2}

Stage~1 produces semantically aligned multimodal shop embeddings.
As shown in Fig.~\ref{fig:stage2-train}, Stage~2 aligns these embeddings with \emph{behavioral} objectives by training a \emph{LLM decoder} to predict the next purchased shop from a retrieved candidate set.
Given a fusion prompt that contains spatiotemporal context, the user’s historical shop sequence, and deliverable candidate shops (all represented by fused multimodal embeddings), the LLM decoder generates an output token sequence whose final token is an \emph{index token} indicating the selected shop in the candidate set.
This generative formulation captures \emph{sequential complementarity} (e.g., ordering milk tea after spicy hotpot) beyond pure content similarity.

\subsubsection{Task and Data Construction}
We formulate Stage~2 as \emph{next-shop prediction under retrieve-and-rank}.
Each training instance is a tuple $(\mathbf{c}, \mathcal{S}_{1:t}, \mathcal{C}, y)$ mined from logs, where
$\mathbf{c}$ denotes spatiotemporal context (e.g., time, location),
$\mathcal{S}_{1:t}=[s_1,\ldots,s_t]$ is the user’s historical shop sequence,
$\mathcal{C}=[c_1,\ldots,c_M]$ is the deliverable candidate set returned by retrieval module,
and $y\in\{1,\ldots,M\}$ is the index of the actually purchased shop within $\mathcal{C}$.
The base objective is to maximize:
\begin{equation}
\max_{\theta}\; \log p_{\theta}\!\left(y \mid \mathbf{c}, \mathcal{S}_{1:t}, \mathcal{C}\right).
\end{equation}

\subsubsection{Fusion Prompt Construction}
We encode $\mathbf{c}$ as descriptive text tokens and prepend them to the sequence.
Each shop (in history and candidates) is represented by a \emph{placeholder token} whose \emph{input embedding} is replaced by the fused multimodal embedding $\mathbf{e}(\cdot)$ learned in Stage~1.
The resulting fusion prompt is:
\begin{equation}
\mathcal{X} = [\mathbf{c}; \mathbf{e}(s_1);\ldots;\mathbf{e}(s_t);\mathbf{e}(c_1);\ldots;\mathbf{e}(c_M)].
\end{equation}

\subsubsection{Output Format and Index Prediction}
Directly predicting shop IDs is infeasible due to the extremely large and dynamic inventory.
We instead predict the \emph{index} within $\mathcal{C}$ using a bounded set of index tokens
$\{\langle \mathrm{index}_1\rangle,\ldots,\langle \mathrm{index}_{M_{\max}}\rangle\}$, where candidate $c_k$ corresponds to $\langle \mathrm{index}_k\rangle$.

\textbf{Reasoning tokens + index token.}
To improve training stability, we let the LLM decoder generate a short sequence of \emph{reasoning tokens} before emitting the final index token.
These reasoning tokens are expected to explain why a candidate shop is chosen given the context and user history:
\begin{equation}
o = [\langle \mathrm{think} \rangle, r_{1:L}, \langle / \mathrm{think} \rangle, \langle \mathrm{index}_k \rangle],
\end{equation}
where $r_{1:L}$ are reasoning tokens and $\langle \mathrm{index}_k \rangle$ indicates the selected candidate index token.

\textbf{Index distribution.}
Let $\mathbf{h}$ denote the hidden state used to predict the final index token. We compute:
\begin{equation}
p(y=k \mid \mathcal{X}) = \mathrm{Softmax}(\mathbf{E}_{\mathrm{idx}}\mathbf{h})_k,\quad k=1,\ldots,M,
\end{equation}
where $\mathbf{E}_{\mathrm{idx}}\in\mathbb{R}^{M_{\max}\times d}$ is a learnable index embedding matrix.
This design matches industrial retrieve-and-rank pipelines, keeps the output vocabulary bounded, and supports long-tail/new shops as long as their embeddings $\mathbf{e}(\cdot)$ are available.

\subsubsection{Backbone and Trainable Parameters}
We adopt Qwen2.5-7B-Instruct as the LLM decoder backbone, since it delivers strong performance among lightweight LLMs.
During Stage~2 training, we update: (i) the full decoder parameters, (ii) the index embedding matrix $\mathbf{E}_{\mathrm{idx}}$, and (iii) the multimodal shop embeddings $\mathbf{e}(\cdot)$.

\subsubsection{SFT Warm-up}
Before RL alignment, we warm-start the LLM decoder with supervised fine-tuning (SFT) to ensure the \emph{reasoning steps} are reasonable and faithful to the ground-truth choice.
For each logged tuple $(\mathbf{c}, \mathcal{S}_{1:t}, \mathcal{C}, y)$, we prompt a much larger teacher LLM (e.g., Qwen-72B) to generate a short rationale $r_{1:L}$ conditioned on the context and the ground-truth decision (e.g., ``it is lunch time and the user has repeatedly ordered from this shop'').
We obtain 1M synthesized $\langle (\mathbf{c}, \mathcal{S}_{1:t}, \mathcal{C}), r_{1:L}, y\rangle$ samples and fine-tune the LLM decoder to generate the output sequence in Eq.~(5), providing a stable and accurate initialization for subsequent GRPO optimization.

\subsubsection{Optimization with GRPO}
Starting from the SFT checkpoint, we further optimize the LLM decoder with GRPO using conversion-based rewards.
Compared with vanilla policy gradients, GRPO uses group-based sampling and within-group advantage normalization, which stabilizes learning under sparse binary rewards.
We additionally apply PPO-style clipping and KL regularization to a reference policy to prevent destructive updates, which is important when jointly updating the decoder and $\mathbf{e}(\cdot)$.

\subsubsection{GRPO Formulation}
We model Stage~2 as RL: the state is the fusion prompt (context, history, and candidate embeddings), the policy is an autoregressive Transformer, and the final decision is represented by the last index token of the generated output sequence.

\textbf{State.}
We define the RL state as the fusion prompt:
\begin{equation}
s \triangleq \mathcal{X} = [\mathbf{c}; \mathbf{e}(s_1);\ldots;\mathbf{e}(s_t);\mathbf{e}(c_1);\ldots;\mathbf{e}(c_M)].
\end{equation}

\textbf{Action/Output.}
The LLM decoder generates an output token sequence $o$; its last token is an index token from
\begin{equation}
\mathcal{A} = \{\langle \mathrm{index}_1\rangle,\ldots,\langle \mathrm{index}_M\rangle\},
\end{equation}
i.e., $\mathrm{last}(o)\in\mathcal{A}$.

\textbf{Policy class.}
The policy $\pi_{\theta}(o\mid s)$ is induced by an autoregressive Transformer decoder (Qwen2.5-7B-Instruct).

\textbf{Reward.}
We consider the purchased shop (conversion/order) as the positive target. The default reward is \emph{binary correctness} based on the final index token:
\begin{equation}
r(s,o)=\mathbb{I}[\mathrm{last}(o)=\langle \mathrm{index}_y\rangle],
\end{equation}
i.e., $r=1$ if the predicted index equals the logged purchased shop, otherwise $r=0$. 
We apply the same scalar reward to all tokens in the generated trajectory under GRPO, including the reasoning tokens.
We do not incorporate utility metrics (e.g., order value/ratings) because they require additional weighting (hyperparameter-sensitive and may misalign objectives) and are sparse/noisy in our logs.

\textbf{GRPO objective.}
For each state $s$, we sample a group of outputs $\{o_{1}, o_{2}, \ldots, o_{G}\}$ from the old policy $\pi_{\theta_{\text{old}}}$ and optimize the policy by maximizing:
\small
    \begin{equation}
    \label{eq:grpo}
    \begin{split}
    &\mathcal{J}_{GRPO}(\theta) = \mathbb{E}\left[s \sim P(S),\{o_i\}_{i=1}^{G} \sim \pi_{\theta_{\text{old}}}(O|s)\right] \\
    &\frac{1}{G} \sum_{i=1}^{G} \frac{1}{|o_i|} \sum_{t=1}^{|o_i|} \bigg\{ \min \bigg[ \frac{\pi_{\theta}(o_{i,t}|s, o_{i,<t})}{\pi_{\theta_{old}}(o_{i,t}|s, o_{i,<t})} \hat{A}_{i,t}, \\
    &\text{clip} \left( \frac{\pi_{\theta}(o_{i,t}|s, o_{i,<t})}{\pi_{\theta_{old}}(o_{i,t}|s, o_{i,<t})}, 1 - \varepsilon, 1 + \varepsilon \right) \hat{A}_{i,t} \bigg]  
    \\
    &- \beta \mathbb{D}_{KL} \left[ \pi_{\theta}\parallel\pi_{\mathrm{ref}} \right]\bigg\},
    \end{split}
    \end{equation}
\normalsize
where $\pi_{\theta}$ and $\pi_{\theta_{\text{old}}}$ are the current and old policy models, and
$\pi_{\mathrm{ref}}$ is a fixed reference policy initialized from the supervised fine-tuning (SFT) checkpoint.
$\varepsilon$ is the clipping hyperparameter and $\beta$ controls KL regularization.
In our setting, each $o_i$ is a short sequence consisting of auxiliary intermediate tokens and a final index token; therefore $|o_i|$ is variable and subject to a maximum generation-length constraint.

\textbf{Advantage computation.}
We use group-relative normalization to reduce variance under sparse rewards:
\begin{equation}
\hat{A}_{i,t}=\widetilde{r}_{i}=\frac{r_i-\mathrm{mean}(\mathbf{r})}{\mathrm{std}(\mathbf{r})},
\end{equation}
where $\mathbf{r}=\{r_1,\ldots,r_G\}$ are rewards of the sampled group for the same state.
We compute a sequence-level advantage $\widetilde{r}_i$ from the terminal reward and broadcast it to all token positions $t$ in $o_i$.
If $\mathrm{std}(\mathbf{r})=0$, we set $\widetilde{r}_i=0$.

\textbf{KL term.}
We regularize the policy with an unbiased, non-negative estimator of the KL divergence between the current policy and the reference policy:
\begin{equation}
\label{eq:kl}
\begin{split}
\mathbb{D}_{\text{KL}} \left[ \pi_\theta \parallel \pi_{\text{ref}} \right] &=
\frac{\pi_{\text{ref}}(o_{i,t}|s, o_{i,<t})}{\pi_\theta(o_{i,t}|s, o_{i,<t})}
- \log \frac{\pi_{\text{ref}}(o_{i,t}|s, o_{i,<t})}{\pi_\theta(o_{i,t}|s, o_{i,<t})} - 1 .
\end{split}
\end{equation}

After Stage~2 training, we export the optimized shop embeddings $\mathbf{e}(\cdot)$ and keep them frozen in Stage~3.
}

\begin{figure}[t]
  \centering
  \includegraphics[width=0.8\linewidth,height=0.25\textheight,keepaspectratio]{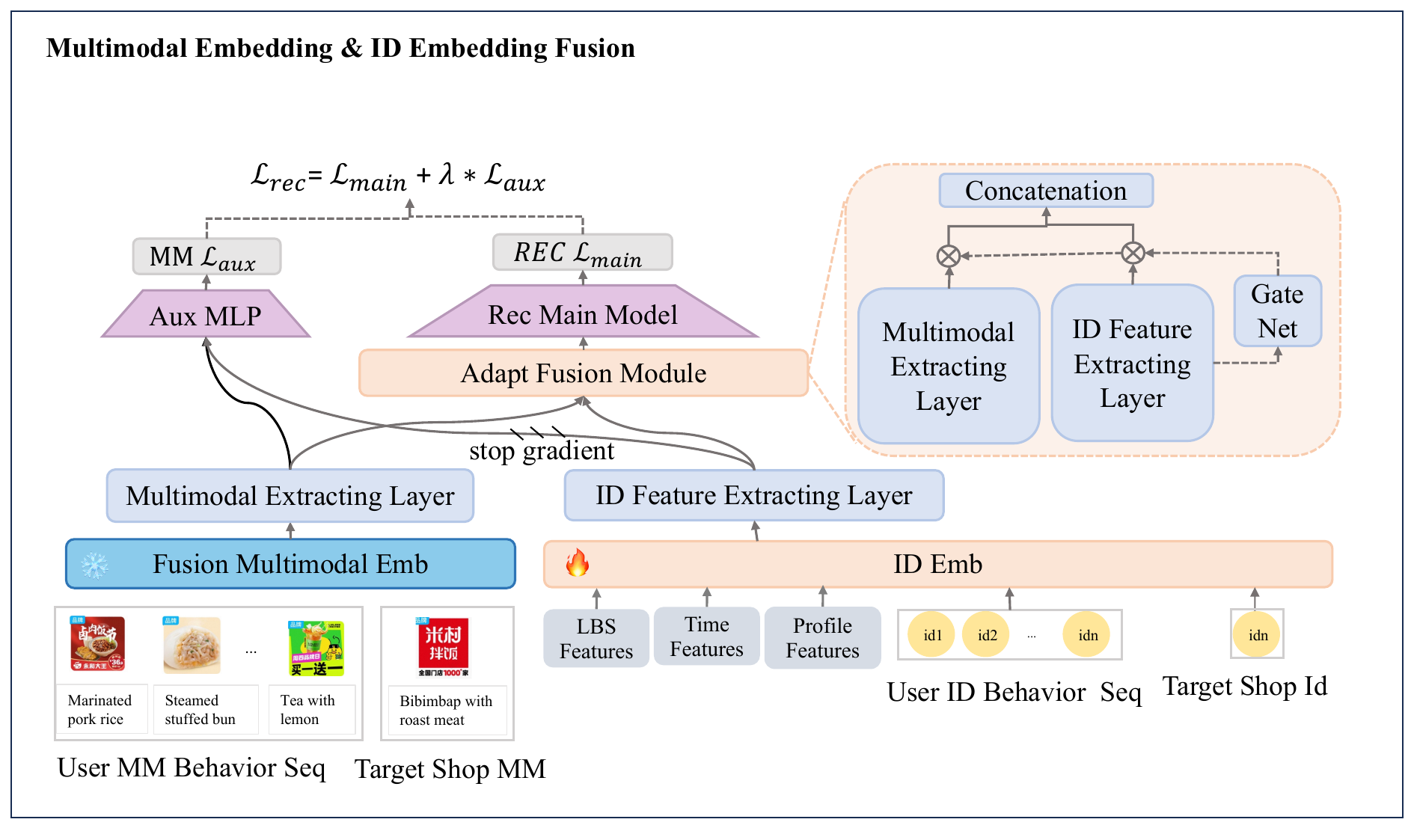}
  \caption{Ranking Model Training Stage of \textbf{GALA}.}
  \label{fig:cotrain}
\end{figure}

\subsection{Multimodal Embedding Fusion for RecSys}
We first outline standard approaches for incorporating multimodal features in recommender systems. Industrial ranking models typically rely on user sequence modeling for personalization. Thus, pre-trained multimodal representations should be integrated through user sequences and target shops. Major approaches process these multimodal features through dedicated interaction layers, analogous to ID-based feature interactions. For instance, \cite{by:25} realized multimodal feature interaction by calculating the similarity between user multimodal sequences and target shop multimodal features; \cite{xr:24} performed bucket processing on similarity scores to enhance the generalization ability of multimodal recommendation. In addition, a fixed-cycle warm-up approach is widely adopted to train ranking models in industrial practices. Most ID-based models are trained on multi-year historical data, resulting in stabilized MLP layer parameters that inherently limit multimodal feature utilization. Crucially, the relative contribution of multimodal representations should adapt to shop characteristics (long-tail vs. popular). For instance, long-tail shops' ID representations often suffer from data sparsity, making their multimodal features comparatively more valuable than those of popular shops. This necessitates an adaptive fusion strategy that dynamically balances ID and multimodal representations to exploit their complementary strengths.

Let the user sequence be denoted as $S_u$ and the target shop as $I_t$, with their multimodal counterparts $S_u^{m}$ and $I_t^{m}$. As illustrated in Fig.~\ref{fig:cotrain}, the adaptive fusion module is positioned after both the multimodal and ID-based representation interaction layers, orchestrating their outputs in the downstream ranking model. The workflow consists of three key steps: First, user behavioral sequences and target shop are processed through their respective ID-based and multimodal feature extractors, yielding ID representation vectors \(h_{id}\) and multimodal representation vectors \(h_{m}\). These two streams undergo interaction modeling within their modalities---ID-based and multimodal interactions---to capture the respective collaborative and content-driven signals. 

To better integrate multimodal information into industrial ranking models, we first design an ID-dependent gating mechanism. This gating network dynamically balances the contributions of ID-based and multimodal representations according to the characteristics of the data, as formulated in Eq.~(\ref{eq5}), where a gating network---whose input is derived from the ID-based representation \(h_{id}\)---computes a dynamic fusion weight \(g\) via a sigmoid-activated linear transformation, and the final fused representation \(h_f\) is computed as a weighted combination of two modalities. Formally, we have:
\begin{equation}
\label{eq5}
\begin{split}
&h_{id}= f_{id}(S_u,I_t),  h_m = f_m(S_u^{m},I_t^{m}),\\
&g= \sigma(W_g \cdot h_{id} + b_g),
h_f = g \cdot h_{id} + (1-g) \cdot h_m ,
\end{split}
\end{equation}
where \( \sigma \) denotes the sigmoid activation function, \( W_g \) and \( b_g \) are trainable parameters, $f_{id}$ is the ID feature interaction function, and $f_m$ is the multimodal feature interaction function. This design enables adaptive weighting between ID-based and multimodal signals based on data characteristics: for data-rich “head” shops with reliable ID embeddings, the gating mechanism prioritizes ID representations, whereas for ``long-tail" or cold-start shops, it shifts emphasis toward multimodal features to mitigate behavioral sparsity. The fused representation is then used as input for the main recommender model, and the primary training objective \(\mathcal{L}_{main}\) is defined as the standard cross-entropy loss for click/conversion prediction as Eq.~(\ref{lmain}). However, due to the long-term accumulation of ID-based features and the stronger learning signals they provide during training, the gating network tends to over-rely on the ID branch. This imbalance leads to gating collapse, where the multimodal pathway is effectively ignored and the advantages of multimodal features are not fully exploited. 
\begin{equation}
\label{lmain}
\mathcal{L}_{\mathrm{main}} = -\frac{1}{N}\sum_{i=1}^{N}\left[y_{i}\log(p_i^{main})+(1 - y_i)\log(1 - p_i^{main})\right],
\end{equation}
where \(N\) is the number of samples in the batch, \(y_{i} \in \{0, 1\}\) represents the ground-truth label (click/conversion) for the \(i\)-th user, and \(p_i^{main}\) is the predicted probability of click/conversion generated by the recommendation main MLP model for the \(i\)-th sample, defined as $p_i^{main} = MLP(h_i^{f})$.

To mitigate this problem, we introduce an auxiliary loss \(\mathcal{L}_{aux}\) as Eq.~(\ref{laux}), which is computed from the concatenated multimodal \(h_{m}\) and ID-based \(h_{id}\) representations (with gradients stopped on the ID branch), directly supervising the multimodal pathway to predict user click/conversion labels. This design ensures gradient flow through the multimodal branch, even when the main loss favors the ID features. 
\begin{equation}
\label{laux}
\mathcal{L}_{aux} = -\frac{1}{N}\sum_{i=1}^{N}\left[y_{i}\log(p_i^{aux})+(1 - y_{i})\log(1 - p_i^{aux})\right],
\end{equation}
where \(p_i^{aux}\) is the predicted probability of click/conversion generated by the recommendation auxiliary MLP model for the \(i\)-th sample, defined as $p_i^{aux} = MLP([\;h_i^{m};\ \mathrm{sg}(h_i^{id})\;])$. Thus, the overall training objective \(\mathcal{L}_{rec}\) can be defined as:
\begin{equation}
\label{lrec}
\mathcal{L}_{rec} = \mathcal{L}_{main} + \lambda \cdot \mathcal{L}_{aux},
\end{equation}
where \( \lambda \) is a key hyperparameter controlling the weight of the auxiliary loss, discussed in the following experiment. 

\section{Experiment}

{\color{revgray}
\subsection{Datasets}

Our experiments are conducted on Taobao Shangou proprietary logs. Our proposed three-stage pipeline requires a specific combination of signals to be evaluated end-to-end: (i) search logs linking user queries to clicked/converted shops/items for Stage~1 query--image--text alignment, (ii) user interaction sequences for Stage~2 next-shop modeling and RL-based refinement, and (iii) recommendation/ranking logs containing exposure behavior together with click/conversion outcomes for Stage~3 ranking/fusion training and evaluation. These signals are standard in real-world industrial search and recommendation systems and can be readily collected in practice.

Specifically, we construct four datasets corresponding to the three stages: for Stage~1, we use (i) 8M verified image--text pairs from merchant uploaded shop/item content (after removing non-core categories), and (ii) 14M query--image--text triplets mined from search-to-purchase logs; for Stage~2, we use (iii) 900M user interaction sequences collected over 3 months; \rev{we further sample $\sim$7M high-quality sequences per week for GRPO training}; for Stage~3, we use (iv) 42B ranking instances constructed from recommendation logs for offline/online evaluation (2-month training window with next-day testing). 

Although we cannot release the raw data due to user privacy and business constraints, we provide more details on the dataset schema, construction/filtering, rules and temporal train/validation/test splits in Table ~\ref{tab:data_construction}.

\begin{table}[t]
\caption{\textbf{Dataset construction summary} for the three-stage pipeline.}
\label{tab:data_construction}
\centering
\scriptsize
\setlength{\tabcolsep}{2pt}
\renewcommand{\arraystretch}{1.0}

\begin{tabular}{@{}p{0.12\columnwidth} p{0.28\columnwidth} p{0.27\columnwidth} p{0.30\columnwidth}@{}}
\toprule
\textbf{Stage} & \textbf{Inputs / Features} & \textbf{Labels (pos/neg)} & \textbf{Split \& Filters} \\
\midrule
Stage~1 &
Per shop: top-10 queries; top-5 purchased items; logo and item images; title/category/desc. &
Pos: (\emph{q}, \emph{img}, \emph{txt}) from search$\rightarrow$purchase.
Neg: in-batch negatives (MoCo-style). &
Random 8:2 train/test.
Filter: exclude non-primary categories (e.g., staples/tableware). \\
\midrule
Stage~2 &
User history; spatiotemporal context; deliverable candidate set. &
Pos: purchased shop in deliverable set.
Neg: other deliverable, non-purchased shops. &
Train: past 3 months exposure/click/order logs.
Test: next-day logs. \\
\midrule
Stage~3 &
ID features (IDs, context, behavior seq, etc.) and frozen MM embeddings. &
Pos: exposed shops with click/order.
Neg: exposed shops without click/order. &
Train: past 2 months ranking logs.
Test: next-day logs. \\
\bottomrule
\end{tabular}
\end{table}
}

\subsection{Baseline Models}
\label{subsec:baselinemodels}
\subsubsection{Compared Representation Methods}

\rev{For multimodal representation learning, we compare against strong and widely adopted vision--language pretraining approaches: ALBEF~\cite{jl:21}, which combines contrastive alignment with multimodal fusion through co-attention and momentum distillation; CNCLIP~\cite{ay:22}, a Chinese CLIP-style model pretrained on large-scale Chinese image--text corpora; GME~\cite{xz:24}, a large multimodal embedder trained with contrastive learning and synthetic data, reporting strong performance on universal multimodal retrieval benchmarks. These methods represent SOTA generic cross-modal alignment. They serve as natural comparators to evaluate whether GALA’s domain-adaptive alignment (Stage 1) and behavior-driven post-training (Stage 2) can produce embeddings that better match recommendation intents.}

\subsubsection{Compared Ranking Models}

\begin{figure}[t]
    \centering
    \subfloat[long-tail bucket vs shop ratio]{
        \includegraphics[width=0.45\linewidth,height=3.2cm,keepaspectratio]{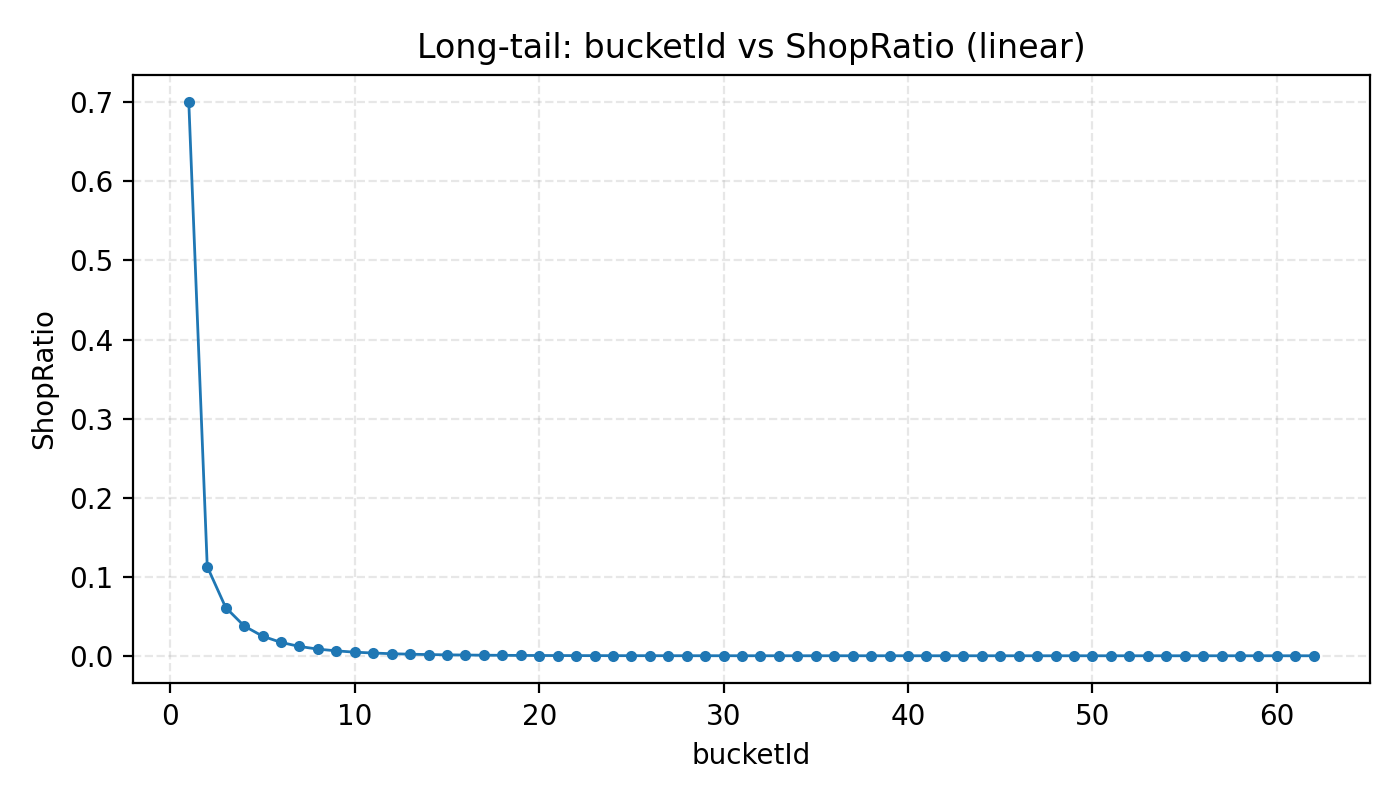}
        \label{fig:linear}
    }\hfill
    \subfloat[coverage rate vs train time]{
        \includegraphics[width=0.45\linewidth,height=2.3cm,keepaspectratio]{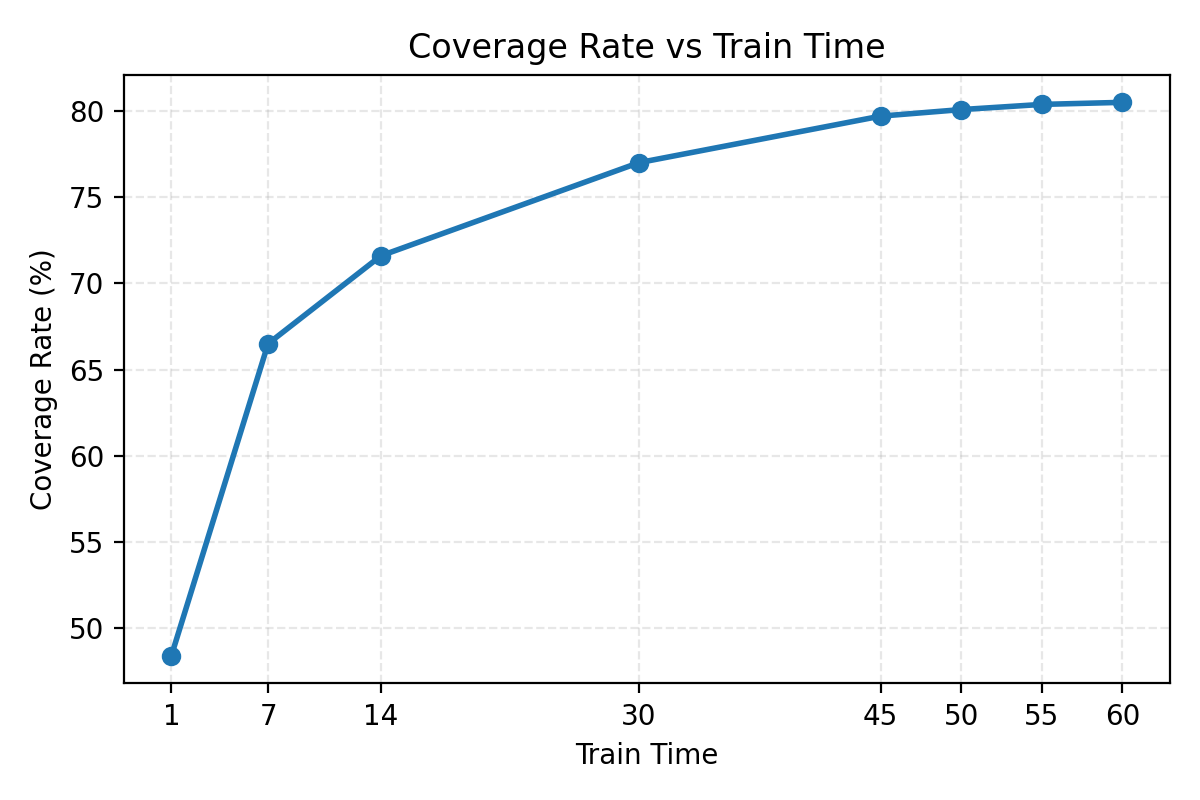}
        \label{fig:log}
    }
    \caption{\textbf{Shop distribution and coverage.}
    (a) Shops are bucketed by their occurrence counts in two months of training data; the shop ratio per bucket shows a pronounced long-tail distribution.
    (b) Coverage of next-day (\(t{+}1\)) candidate shops by training data aggregated over the past \(k\) days (\(t{-}1\) to \(t{-}60\)); coverage increases with more history and plateaus at \(\sim 80\%\).}
    \label{fig:longtail}
\end{figure}

{\color{revgray}
Our baseline choices are driven by the production deployment constraint in Taobao Shangou: the online shop ranker must satisfy strict millisecond-level latency and high-throughput serving. Therefore, the system follows the standard industrial paradigm of \emph{offline multimodal embedding computation, online KV lookup, a lightweight ranker}, i.e., \emph{frozen-embedding deployment} at serving time. Under this setting, the most relevant comparisons are (i) how multimodal representations are injected/fused into the ranker given frozen embeddings, and (ii) how the quality of the frozen multimodal embeddings affects ranking when using the same fusion mechanism. 

Although end-to-end multimodal rankers that jointly update encoders with ranking objectives are a strong algorithmic alternative, they are difficult to deploy under our production constraints. In particular, such methods often require per-request multimodal encoding or frequent encoder refresh, which is prohibitive under strict millisecond-level latency. Moreover, many recent end-to-end designs rely on cached/memory-bank item representations to reduce cost, which introduces a long-tail \emph{coverage} issue that is particularly severe in our fast-evolving shop inventory (Fig.~\ref{fig:linear}). As shown in Fig.~\ref{fig:log}, even under a relaxed criterion (``at least one conversion observed''), aggregating training data from the past $k$ days ($k{=}1$ to $60$) covers at most $\sim$80\% of $t{+}1$ candidate shops, leaving a non-trivial fraction without up-to-date representations. In contrast, our frozen-embedding pipeline supports daily offline incremental inference for newly added/updated shops and achieves near-complete coverage (99.975\% in our measurements) while meeting online latency constraints. Therefore, we focus on deployable baselines under the frozen-embedding serving assumption.

\textbf{SOTA ways of injecting multimodal representations into the ranker.} We compare representative and deployable multimodal ranking/fusion methods that operate on frozen embeddings:
\begin{itemize}
    \item MMREC: a typical two-stage approach that directly concatenates/injects multimodal content embeddings into the ranking model for CTR/CVR prediction.
    \item SimTier: incorporates multimodal signals via similarity computation (e.g., user--item/content similarity features and subsequent processing), and feeds these similarity-based features into the ranker.
    \item AlignRec: a two-stage alignment-based method that first performs content normalization/alignment (e.g., image--text alignment) and then uses contrastive alignment between multimodal and ID representations before integrating them into the ranker.
    \item LUM: a three-stage design that first aligns image--text content, then performs generative multimodal user-sequence modeling to obtain a frozen user multimodal preference representation, which is finally injected into the ranker.
\end{itemize}
These baselines cover major industrially feasible fusion paradigms (direct injection, similarity-based features, representation alignment, and generative user modeling) under the same frozen-embedding serving assumption.

\textbf{Ablations on multimodal representations under our adaptive fusion framework.} To isolate the impact of multimodal representation quality, we further conduct ablations by combining different content embeddings with our proposed adaptive fusion of ID and content representations:
\begin{itemize}
    \item GALA-gme\_emb: uses frozen multimodal embeddings extracted from an industry SOTA multimodal foundation model (GME) and applies our adaptive ID--content fusion in the ranker. This tests the effect of a strong off-the-shelf embedder under the same fusion module.
    \item GALA-image\_emb / GALA-text\_emb: use the single-modality image-only or text-only embeddings trained by our representation learning pipeline, combined with the same adaptive fusion module, to quantify each modality’s contribution.
    \item GALA: uses our learned multimodal fused content embedding together with the adaptive fusion module, representing the full model.
\end{itemize}
}

\subsection{Experiment Settings}
%\subsubsection{Experiment Settings}
\label{subsec:experimentsettings}
We implement Domain-Adaptive Cross-Modal Embedding Alignment based on the GME framework, utilizing its unified joint encoder as the backbone. Final shop and item representations are obtained from the output embedding of the last token. For cross-modal representation learning, we systematically compare two fusion strategies: (i) a baseline multi-layer perceptron (MLP), and (ii) a T5-style cross-attention fusion module that explicitly captures bidirectional cross-modal interactions.

All experiments are conducted with a global batch size of $24,576$ (using 24 gradient accumulation steps and a per-device batch size of $8$). Optimization settings include a base learning rate of $2\times10^{-5}$, weight decay of $0.1$, and a learning rate schedule with $0.05$ linear warmup followed by cosine decay. The contrastive loss for image-to-text alignment follows Eq.~(\ref{eq1}), where the temperature parameter $\tau$ is fixed at $2.0$ throughout training. The overall training objective combines three loss components as in Eq.~(\ref{eq2}), with weighting coefficients set to $w_1=0.3$, $w_2=0.3$, and $w_3=0.4$.

For the next-shop prediction stage, the learning rate is set to $5\times10^{-5}$. We adopt \textsc{GRPO} with clipping ($\epsilon=0.2$) and KL regularization ($\beta=0.001$) for stability, and cap the policy model's maximum generation length at 2048 tokens. The ranking hyperparameter $\lambda$ is set to 0.01, which empirically facilitates effective cross-modal interactions between visual features and textual descriptors.

\subsection{Evaluation Metrics}
\label{subsec:evaluationmetrics}
\subsubsection{Evaluating Representation Method}
We systematically assess embedding quality with three retrieval tasks: \textbf{1) Item Intention Task}: Retrieves items for queries with explicit names (e.g., ``burgers''), testing fine-grained semantic capture; \textbf{2) Shop Intention Task}: Identifies correct shops from brand-specific queries (e.g., ``KFC''), evaluating brand and establishment-level understanding; \textbf{3) Composite Intention Task}: Handles queries combining brand and item (e.g., ``KFC burgers''), requiring multi-level semantic integration.
We use Recall@K as the evaluation metric. Recall@K is defined as: 
\begin{equation}
\label{eq:recall}
\mathrm{Recall@}K = \dfrac{ \left| \mathrm{Top}\text{-}K \cap \mathcal{R} \right| }{ \left| \mathcal{R} \right| },
\end{equation}
where $\mathcal{R}$ is the set of ground-truth relevant entities per query. Positive samples are derived from user search queries that led to actual purchases, as captured in production logs. We compute the Recall@K for $K \in \{1, 5, 10, 20\}$ and use their average as final evaluation metrics to reflect different retrieval depths.

\subsubsection{Evaluating Ranking Models}
We evaluate ranking performance using AUC~\cite{ap:97} and PCOC~\cite{10.1145/2648584.2648589}.
AUC is the probability that a randomly sampled positive instance is ranked above a randomly sampled negative one.
PCOC is defined as
\begin{equation}
\label{eq:pcoc}
\mathrm{PCOC}=\frac{\mathbb{E}[\hat{p}]}{\mathbb{E}[y]},
\end{equation}
where the expectation is taken over the evaluation set, $\hat{p}$ is the predicted CTR/CVR, and $y\in\{0,1\}$ is the click/conversion label (optimal $\approx 1$; $<1$ underestimates and $>1$ overestimates).

\subsection{Offline Performance}
\label{subsec:offlineperformance}

\subsubsection{Different Representation Learning Strategies}
In this subsection, we evaluate \textbf{GALA} from two perspectives: (i) quantitative comparison on recall performance against SOTA embedding models, and (ii) qualitative visualization of its two-stage alignment (Stage 1--2) to illustrate how multimodal representations evolve.

\begin{table}[htbp]
\caption{Average Recall@K Scores Across Retrieval Tasks}
\begin{center}
\begin{tabular}{c|c|c|c|c|c}
\hline
\textbf{Model} & \textbf{Dim} & \textbf{Item} & \textbf{Shop} & \textbf{Comp} & \textbf{Overall} \\
\hline
ALBEF          & 128  & 0.724 & 0.107 & 0.791 & 0.541 \\
CNCLIP         & 768  & 0.691 & 0.471 & 0.782 & 0.648 \\
GME            & 1536 & 0.788 & 0.869 & 0.841 & 0.843 \\
\hline
GALA (MLP)     & 128  & 0.841 & 0.877 & 0.851 & 0.856 \\
GALA (T5)      & 128  & 0.844 & 0.887 & 0.848 & 0.860 \\
GALA (T5+GRPO) & 128  & \textbf{0.866} & \textbf{0.896} & \textbf{0.868} & \textbf{0.877} \\
\hline
\end{tabular}
\label{tab:recall}
\end{center}
\end{table}

\textbf{Performance Comparison of Different Representation Learning:} Table~\ref{tab:recall} presents the retrieval performance across diverse search intentions. For brevity, the full results use abbreviated names: ``Item'', ``Shop'', and ``Comp'' denote \textit{Item Intention}, \textit{Shop Intention}, and \textit{Composite Intention} retrieval tasks, respectively. Model variants of \textbf{GALA} are shortened: ``(MLP)'' stands for MLP-fusion, ``(T5)'' for T5-fusion, and ``(T5+GRPO)'' includes GRPO optimization. Dimensions (Dim) are included in parentheses. Our framework consistently outperforms existing approaches, with the T5-Fusion+GRPO variant establishing new SOTA results (Recall@K = 0.877), achieving a 3.4\% absolute improvement over the strongest baseline (GME). Notably, these advancements are achieved while maintaining compact 128-dimensional embeddings (vs. 1536D in GME), demonstrating the efficiency of our fusion mechanism. The GRPO module in \textbf{GALA} contributes large gains (+1.7\% over T5-Fusion, +2.1\% over MLP-Fusion), validating its effectiveness in multimodal alignment. It is worth noting that \textbf{GME} is evaluated using its released configuration with a dimensionality of 1536, without any modification. In contrast, \textbf{GALA} adopts a compact 128-dimensional embedding, which is more suitable for production deployment environments by reducing latency, memory footprint, and overall serving cost.

Analysis of baselines reveals critical limitations: ALBEF shows severe imbalance (0.107 recall for shop intention), while CNCLIP's high-dimensional design yields suboptimal performance (0.471). These underscore the importance of intention-aware fusion strategies in retrieval tasks.

\begin{figure}[t]
    \centering
    \begin{minipage}{\linewidth}
        \centering
        \begin{subfigure}[t]{0.48\linewidth}
            \centering
            \includegraphics[width=\linewidth]{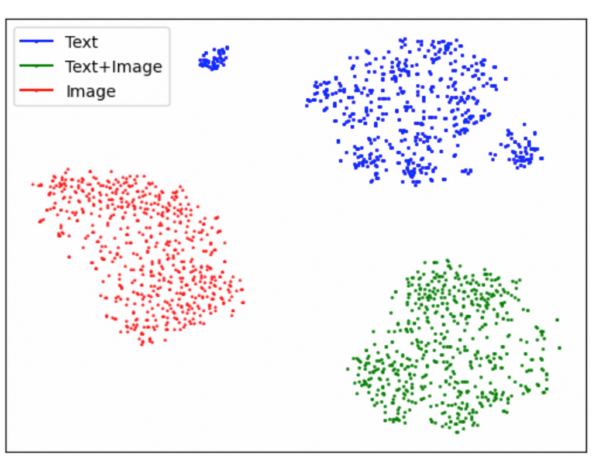}
            \caption{Before training: multimodal features (Text, Text+Image, Image) occupy distinct and disjoint regions in the embedding space.}
            \label{fig:tsne_before}
        \end{subfigure}\hfill
        \begin{subfigure}[t]{0.48\linewidth}
            \centering
            \includegraphics[width=\linewidth]{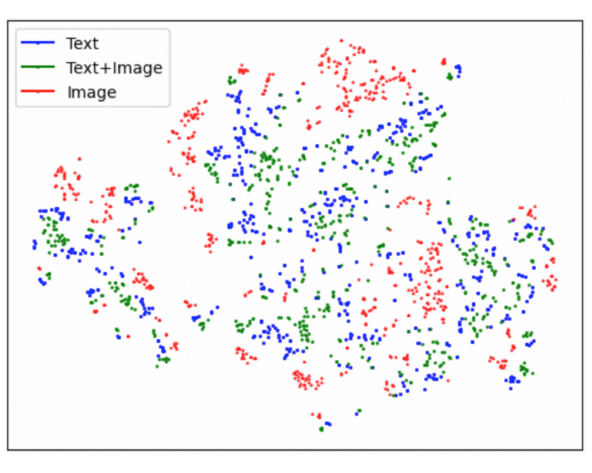}
            \caption{After training: features from different modalities exhibit significant overlap, indicating improved cross-modal coherence.}
            \label{fig:tsne_after}
        \end{subfigure}
    \end{minipage}
    \caption{t-SNE visualization of multimodal feature distributions before and after Stage~1 \textit{Domain-Adaptive Alignment}.}
    \label{fig:tsne}
\end{figure}

\begin{figure}[t]
    \centering

    % ===== Row 1 =====
    \begin{minipage}{0.95\linewidth}
        \centering
        \begin{subfigure}[t]{0.48\linewidth}
            \centering
            \includegraphics[width=\linewidth]{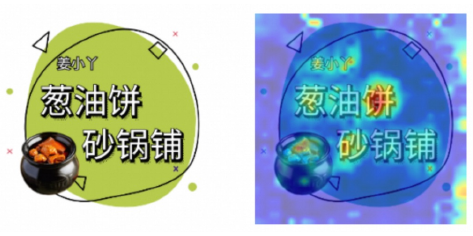}
            \caption{Attention for query keyword: ``Pancake''.}
            \label{fig:query_a}
        \end{subfigure}\hfill
        \begin{subfigure}[t]{0.48\linewidth}
            \centering
            \includegraphics[width=\linewidth]{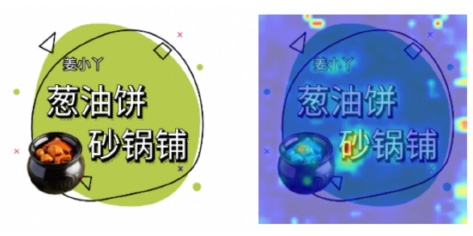}
            \caption{Attention for query keyword: ``Clay Pot''.}
            \label{fig:query_b}
        \end{subfigure}
    \end{minipage}

    \vspace{0.5em}

    % ===== Row 2 =====
    \begin{minipage}{0.95\linewidth}
        \centering
        \begin{subfigure}[t]{0.48\linewidth}
            \centering
            \includegraphics[width=\linewidth]{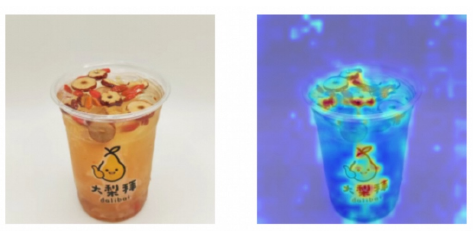}
            \caption{Attention for query keyword: ``Pear Soup''.}
            \label{fig:query_c}
        \end{subfigure}\hfill
        \begin{subfigure}[t]{0.48\linewidth}
            \centering
            \includegraphics[width=\linewidth]{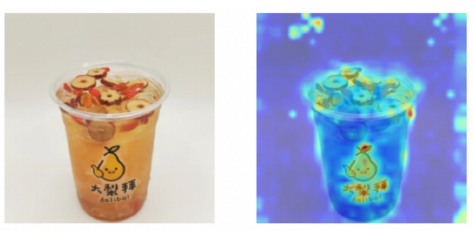}
            \caption{Attention for query keyword: ``Cup''.}
            \label{fig:query_d}
        \end{subfigure}
    \end{minipage}

    \caption{Visualization of model attention under different query keywords.
Each panel overlays the attention heatmap on the original image,
highlighting query-specific semantic focus regions.}
    \label{fig:query_attn}
\end{figure}

\textbf{Analysis of Domain-Adaptive Alignment.} Following Stage~1 domain-adaptive alignment, the t-SNE visualization (Fig.~\ref{fig:tsne}) reveals that multimodal features---originally residing in disjoint vector spaces---are now closely integrated. Specifically, image and text embeddings, which initially formed distinct clusters, exhibit substantial overlap after alignment, indicating significantly enhanced cross-modal coherence. Moreover, attention response analysis under diverse query inputs (Fig.~\ref{fig:query_attn}) demonstrates that the model dynamically shifts its attention focus to the most relevant modality-specific regions as query semantics evolve, reflecting improved sensitivity to user intent during retrieval.

\begin{figure*}[t]
    \centering
    % ==== 图 a ====
    \begin{subfigure}[t]{0.23\textwidth}
        \centering
        \includegraphics[width=0.8\linewidth]{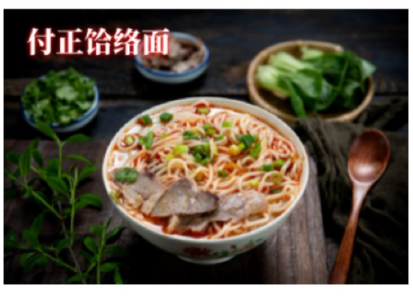} % 原图路径
        \caption{Original image}
        \label{fig:attn_a}
    \end{subfigure}
    \hfill
    % ==== 图 b ====
    \begin{subfigure}[t]{0.23\textwidth}
        \centering
        \includegraphics[width=0.8\linewidth]{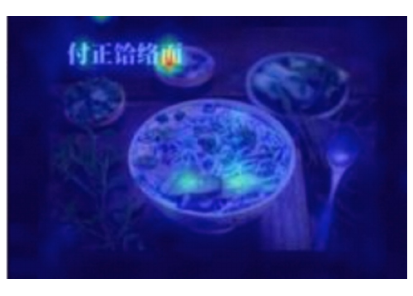} % GME Attention
        \caption{GME attention visualization}
        \label{fig:attn_b}
    \end{subfigure}
    \hfill
    % ==== 图 c ====
    \begin{subfigure}[t]{0.23\textwidth}
        \centering
        \includegraphics[width=0.8\linewidth]{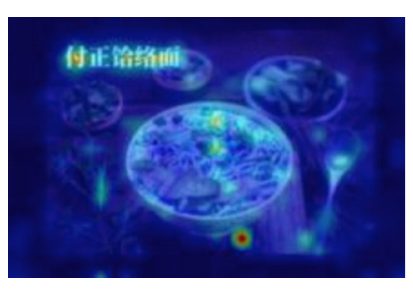} % GALA(T5) Attention
        \caption{GALA(T5) attention visualization}
        \label{fig:attn_c}
    \end{subfigure}
    \hfill
    % ==== 图 d ====
    \begin{subfigure}[t]{0.23\textwidth}
        \centering
        \includegraphics[width=0.8\linewidth]{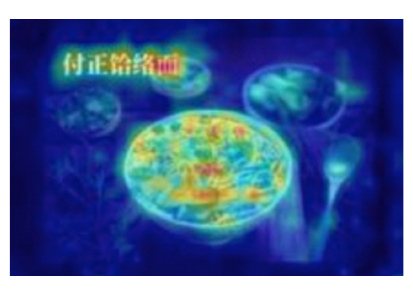} % GALA(T5+GRPO) Attention
        \caption{GALA(T5+GRPO) attention visualization}
        \label{fig:attn_d}
    \end{subfigure}
    
    \caption{Comparison of attention maps under different methods for the given query "qia luo noodles". The visualization shows that GALA variants progressively enhance focus on query-relevant regions compared to the baseline. Notably, the multimodal representations produced by GALA (T5+GRPO) more faithfully capture the key content areas aligned with the user's core attentional focus.}
    \label{fig:attn_comparison}
\end{figure*}

\textbf{Analysis of Generative User-Behavior Alignment.} In Stage~2, generative alignment grounded in historical user behavior logs further refines cross-modal interactions. Post-training attention heatmaps (Fig.~\ref{fig:attn_comparison}) show that GALA variants progressively concentrate attention on query-relevant regions, outperforming the baseline. Notably, the multimodal representations generated by GALA (T5+GRPO) more accurately capture key content areas that align with users’ core attentional focus. Together, these visualization results validate \textbf{GALA}’s dual capability: achieving seamless integration of cross-modal embedding spaces and enabling intent-aware attention modulation.

\subsubsection{Different Ranking Model Training Strategies}

We evaluate \textbf{GALA} through comprehensive experiments to: (i) benchmark ranking against SOTA baselines, (ii) quantify each modality's contribution via ablations, and (iii) analyze sensitivity to key hyperparameters and the efficacy of adaptive fusion.

\begin{table}[htbp]
\caption{Ranking Performance of Different Methods}
\begin{center}
\begin{tabular}{c|c|c|c|c}
\hline
\multirow{2}{*}{\textbf{Model}} & \multicolumn{2}{c|}{\textbf{AUC}} & \multicolumn{2}{c}{\textbf{PCOC}} \\
\cline{2-5}
 & \textbf{CTR} & \textbf{CVR} & \textbf{CTR} & \textbf{CVR} \\
\hline
MMREC             & 0.7242 & 0.8158 & 1.0435 & 1.0832 \\
SimTier           & 0.7237 & 0.8155 & 1.0338 & 1.0649 \\
AlignRec          & 0.7243 & 0.8162 & 1.0321 & 1.0521 \\
LUM               & 0.7251 & 0.8171 & 1.0302 & 1.0625 \\
\hline
GALA-gme\_emb     & 0.7251 & 0.8165 & 1.0388 & 1.1027 \\
GALA-image\_emb   & 0.7250 & 0.8167 & 1.0442 & 1.0982 \\
GALA-text\_emb    & 0.7253 & 0.8168 & 1.0484 & 1.0912 \\
\hline
GALA              & \textbf{0.7263} & \textbf{0.8193} & \textbf{1.0212} & \textbf{1.0276} \\
\hline
\end{tabular}
\label{tab:rank_table1}
\end{center}
\end{table}

\textbf{Performance Comparison of Different Ranking Models.} Table \ref{tab:rank_table1} presents a comparative evaluation of \textbf{GALA} against mainstream ranking models (MMREC, SimTier, AlignRec, LUM) and its ablated variants (with different modal components removed). The key findings are: 1) \textbf{GALA} achieves the best performance across all metrics---CTR-AUC (0.7263), CVR-AUC (0.8193), CTR-PCOC (1.0212), and CVR-PCOC (1.0276)---verifying its dual advantages in ranking and prediction accuracies. 2) Importance of multimodal fusion: the ablated \textbf{GALA} variants\footnote{GALA-gme\_emb replaces Fusion\_emb with the GME backbone's output embedding (gme\_emb); GALA-image\_emb uses only image modality features (image\_emb); and GALA-text\_emb employs only text modality features (text\_emb).} consistently underperform the complete model, demonstrating the complementary value of each modality component. For example, after removing text embeddings (text\_emb), CVR-PCOC moves further away from 1. 

\begin{table*}[htbp]
\caption{Ranking Performance Across Different Shop Stratifications}
\begin{center}
\begin{tabular}{c|c|c|c|c|c|c|c|c}
\hline
\multirow{2}{*}{\textbf{Shop Stratification}} & \multicolumn{2}{c|}{\textbf{CTR AUC}} & \multicolumn{2}{c|}{\textbf{CVR AUC}} & \multicolumn{2}{c|}{\textbf{CTR PCOC}} & \multicolumn{2}{c}{\textbf{CVR PCOC}} \\
\cline{2-9}
 & \textbf{Online Base} & \textbf{GALA} & \textbf{Online Base} & \textbf{GALA} & \textbf{Online Base} & \textbf{GALA} & \textbf{Online Base} & \textbf{GALA} \\
\hline
L6       & 0.7230 & \textbf{0.7243} & 0.8140 & 0.8140 & 1.1132 & \textbf{1.0958} & 1.2024 & \textbf{1.1754} \\
L5       & 0.7234 & \textbf{0.7258} & 0.8239 & \textbf{0.8271} & 1.1072 & \textbf{1.0886} & 1.1891 & \textbf{1.1600} \\
L4       & 0.7302 & \textbf{0.7316} & 0.8302 & \textbf{0.8333} & 1.0904 & \textbf{1.0711} & 1.1704 & \textbf{1.1352} \\
L3       & 0.7410 & \textbf{0.7434} & 0.8349 & 0.8349 & 1.0695 & \textbf{1.0481} & 1.1340 & \textbf{1.0919} \\
L2       & 0.7545 & \textbf{0.7567} & \textbf{0.8382} & 0.8377 & 1.0530 & \textbf{1.0299} & 1.1005 & \textbf{1.0520} \\
L1       & 0.7706 & \textbf{0.7739} & 0.8379 & \textbf{0.8401} & 1.0354 & \textbf{1.0096} & 1.0560 & \textbf{1.0007} \\
\hline
Overall  & 0.7240 & \textbf{0.7263} & 0.8156 & \textbf{0.8193} & 1.0409 & \textbf{1.0212} & 1.0608 & \textbf{1.0276} \\
\hline
\end{tabular}
\label{tab:rank_table2}
\end{center}
\end{table*}

\textbf{Ranking Performance Across Different Shop Stratifications.} Table \ref{tab:rank_table2} compares the performance of \textbf{GALA} with the online baseline model by shop stratification (L1 to L6, where L1 is usually tail shops and L6 is head shops). The results show: Performance improvement across all stratifications: \textbf{GALA} is comprehensively superior to the baseline in overall metrics, with CTR-AUC increasing by 0.0023, CVR-AUC increasing by 0.0037, CTR-PCOC optimized from 1.0409 to 1.0212 (closer to 1), and CVR-PCOC optimized from 1.0608 to 1.0276, verifying its adaptability to all shops.

\begin{table}[htbp]
\caption{Ranking Performance Across Different Auxiliary Loss Weights}
\begin{center}
\begin{tabular}{c|c|c|c|c}
\hline
& \multicolumn{2}{c|}{\textbf{AUC}} & \multicolumn{2}{c}{\textbf{PCOC}} \\
\cline{2-5}
& \textbf{CTR} & \textbf{CVR} & \textbf{CTR} & \textbf{CVR} \\
\hline
$\lambda=0.01$ & \textbf{0.7263} & \textbf{0.8193} & 1.0212 & \textbf{1.0276} \\
$\lambda=0.1$  & 0.7261 & 0.8187 & 1.0245 & 1.0400 \\
$\lambda=1$    & 0.7150 & 0.8102 & \textbf{1.0196} & 1.0484 \\
\hline
\end{tabular}
\label{tab:rank_table3}
\end{center}
\end{table}

\textbf{Hyperparameter Analysis.} Table \ref{tab:rank_table3} reveals two key findings: 1) \textbf{GALA} achieves the best performance across all metrics when $\lambda=0.01$; 2) As $\lambda$ increases (0.01→0.1→1), the AUC metric decreases significantly (e.g., CVR-AUC drops from 0.8193 to 0.8102), and the PCOC also deviates (CVR-PCOC rises from 1.0276 to 1.0484). This demonstrates the need for careful balance in auxiliary loss weighting, since excessive $\lambda$ values induce optimization conflicts between the primary ranking objective and auxiliary tasks. 

\begin{figure}[ht]
  \centering
  \includegraphics[width=0.9\linewidth, height = 0.45\linewidth]{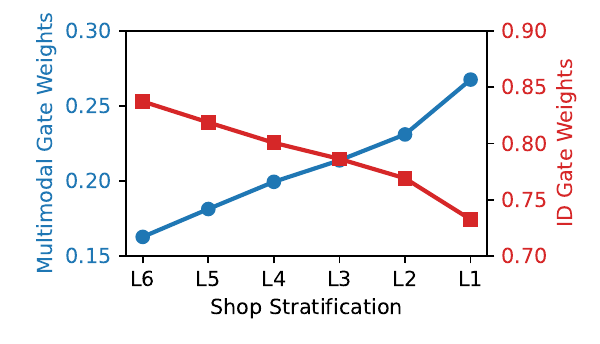}
  \caption{Contributions of Multimodal/ID Gate Weights Across Different Shop Stratifications.}
  \label {fig:adapt_weights}
\end{figure}

\textbf{Analysis of Multimodal Adaptive Weights.} Fig.~\ref{fig:adapt_weights} shows the adaptive weight distribution of multimodal embeddings in \textbf{GALA}, revealing the following conclusions: From head L6 to tail L1, the contribution weight of multimodal representations gradually increases. That is, for popular IDs (with rich data and sufficient training), the model assigns higher weights to ID representations, and the contribution of multimodal representations is relatively reduced; In the scenario of medium and long-tail IDs (with limited training data), the model increases the weight proportion of multimodal representations to make up for the deficiency of ID representations. This adaptive mechanism enables dynamic optimization of representation fusion based on input data characteristics. In other words, \textbf{GALA} can adaptively allocate modal weights based on the input data, thereby enhancing the model's multimodal information utilization efficiency.

\begin{figure}[t]
  \centering
  \includegraphics[width=\linewidth]{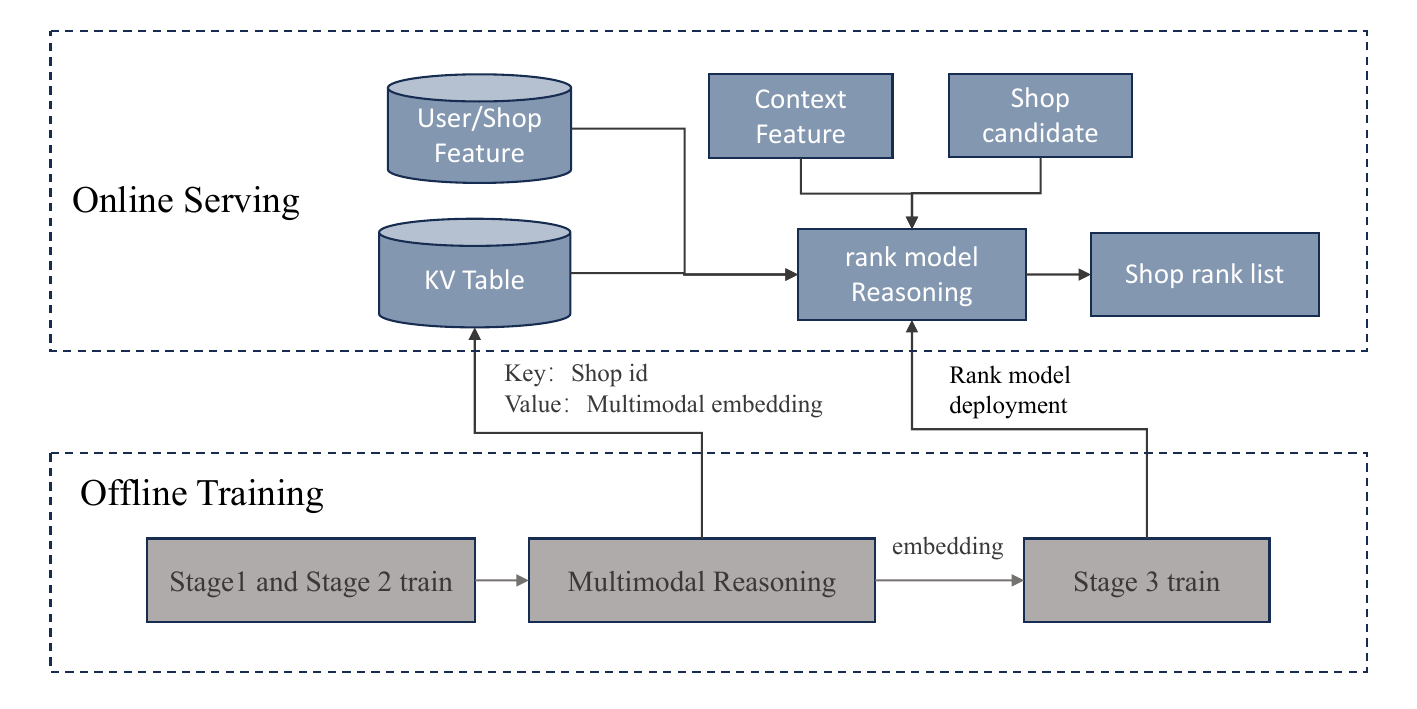}
  \caption{Offline Training and Offline Deployment of GALA.}
  \label {fig:online-offline}
\end{figure}

{\color{revgray}
\subsection{Complexity and Overhead Analysis}
\label{subsec:complexity_overhead}
We quantify the incremental overhead introduced by \textbf{GALA} along three axes: (i) offline training cost, (ii) offline incremental embedding inference cost, and (iii) online serving cost (latency/throughput/memory).

\subsubsection{Sources of additional complexity}
Compared with the baseline multimodal ranker, GALA introduces: (i) two additional offline stages (Stage~1 representation pretraining; Stage~2 GRPO-based behavior alignment); (ii) daily incremental embedding inference for newly added/updated content; (iii) an online KV lookup to fetch frozen multimodal embeddings and fuse them with existing ranking features.

\subsubsection{Offline training overhead}
GALA shifts most complexity offline. The frequency and sample size of incremental training are different in different training stages offline. Stage~1 representation pretraining is executed weekly on $\sim$5M query--image--text pairs and takes $\sim$18 hours. Stage~2 GRPO-based behavior alignment is also executed weekly on $\sim$7M user behavior sequences and takes $\sim$24 hours. Stage~3 fine-tunes the ranking model daily on $\sim$0.7B samples and takes $\sim$3 hours. In Stage~3, the multimodal embeddings from Stages~1--2 are used as fixed (frozen) representations and are not updated, leading to an additional training overhead of $\sim$10\% compared with the baseline multimodal ranker.

\subsubsection{Offline incremental embedding inference}
For newly added/updated content, we perform daily incremental embedding inference for $\sim$150K images and $\sim$76K texts, with an end-to-end inference time of $\sim$30 minutes.

\subsubsection{Online serving overhead}
During online inference, the pretrained multimodal embeddings are stored in a KV table, where the key is the shop identifier
and the value is its multimodal representation. For each candidate shop, the ranker retrieves the corresponding multimodal embedding via KV lookup and fuses it with user/shop/context features to estimate CTR/CVR scores. We benchmark online inference on a production-like machine (96 CPU cores and PPU610), reporting the average over two machines to reduce variance; throughput is measured under a 55\% online accelerator safe-utilization cap. Under this setting, GALA increases the P99 ranking latency from 15.9\,ms to 17.6\,ms (i.e., +1.7\,ms), which is within our production latency budget, with negligible impact on serving throughput and memory usage.

\subsection{Online Effect}
\label{subsec:Online effect}
Since November 2024, \textbf{GALA} has been incorporated into the shop recommender system at Taobao Shangou. 
We conduct a randomized traffic-split online A/B test and compute statistical significance based on day-level uplifts to avoid overstating significance under massive user counts. Most notably, the order volume increases by +0.55\% (95\% CI: [0.342\%, 0.756\%], $p < 0.01$), demonstrating a statistically significant business gain. Meanwhile, GALA improves shop exposure width by 0.5\% overall, with larger gains during peak dining hours (morning +1.78\%, evening +1.05\%), indicating increased visibility for long-tail shops and improved recommendation novelty.
}

\section{Conclusion}
In this paper, we present \textbf{GALA}, a framework designed to learn adaptive multimodal representations for complex food delivery platforms. \textbf{GALA} addresses the mismatch between static content-aligned embeddings and dynamic behavioral goals (CTR/CVR) by introducing a generative reinforcement learning (RL) alignment stage. In this stage, we construct a multimodal behavior-alignment dataset from historical user interactions and optimize it directly with conversion-based rewards. This ensures the pretraining distribution is inherently aligned with the downstream ranking objective. The \textbf{GALA} framework is built on two core components to maximize alignment efficacy: (i) Stage 1 leverages query--image--text triplets extracted from search logs to inject user intent early into the training pipeline, narrowing the semantic gap prior to RL-based optimization; and (ii) an adaptive gating mechanism with auxiliary supervision is employed to preserve informative multimodal signals under ID-dominated training regimes. Through this design, \textbf{GALA} learns fine-grained content representations that remain semantically consistent across modalities, capture dynamic user intents via GRPO-optimized behavior modeling, and enable robust feature integration through adaptive fusion of multimodal and ID-based representations. Comprehensive offline experiments demonstrate that \textbf{GALA} consistently outperforms strong baselines in retrieval and ranking performance, while online A/B testing on Taobao Shangou platform confirms its practical impact, with deployment leading to a 0.55\% increase in purchase volume. These results validate that the proposed generative alignment framework effectively bridges the gap between content-level pretraining and behavior-driven downstream objectives, offering a deployable solution for real-world multimodal recommendation scenarios. We plan to explore deployable end-to-end training under long-tail coverage constraints. A key direction is to explicitly identify long-tail shops (e.g., by exposure/conversion frequency and content freshness) and design \emph{differentiated} representation and optimization strategies for tail vs.\ head shops. For \emph{head} shops with abundant interactions, we will prioritize behavior-driven optimization that fine-tunes their multimodal representations with stronger ranking/RL signals to capture subtle preference shifts. For \emph{long-tail} shops with sparse behavior, we will adopt a coverage-oriented multimodal strategy that relies more on content supervision and robustness to missing/low-quality modalities, together with freshness-aware mechanisms (e.g., staleness modeling and fallback embeddings) to guarantee reliable representations at serving time. Such a differentiated multimodal optimization scheme for head vs.\ tail shops can better balance effectiveness and production constraints without sacrificing long-tail coverage.

\section*{AI-Generated Content Acknowledgement}
No AI-generated content was used in the creation of this manuscript.

\begingroup
\setlength{\itemsep}{0pt}
\setlength{\parskip}{0pt}
\setlength{\parsep}{0pt}
\renewcommand{\baselinestretch}{0.97}\normalsize
\bibliographystyle{IEEEtran}
\bibliography{icde2026}
\endgroup

\end{document}